\documentstyle[epsfig,12pt]{article}
\def\slashchar#1{\setbox0=\hbox{$#1$}           
   \dimen0=\wd0                                 
   \setbox1=\hbox{/} \dimen1=\wd1               
   \ifdim\dimen0>\dimen1                        
      \rlap{\hbox to \dimen0{\hfil/\hfil}}      
      #1                                        
   \else                                        
      \rlap{\hbox to \dimen1{\hfil$#1$\hfil}}   
      /                                         
   \fi}                                         %
\newcommand{\newc}{\newcommand}
\newc{\gsim}{\lower.7ex\hbox{$\;\stackrel{\textstyle>}{\sim}\;$}}
\newc{\lsim}{\lower.7ex\hbox{$\;\stackrel{\textstyle<}{\sim}\;$}}
\begin{document}
\newcommand{\mtildet}{M_{\tilde t_1}}
\def\sps{superpartners\ }
\def\sp{superpartner\ }
\def\ss{supersymmetry\ }
\def\sc{supersymmetric\ }
\def\sm{Standard Model}
\title{TESTS AND IMPLICATIONS OF INCREASING EVIDENCE FOR
SUPERPARTNERS\footnote{Invited talk at XXXII Rencontres des Moriond, Les
Arcs, March 1997}}
\author{Gordon L. Kane\\
Randall Lab of Physics,
University of
Michigan,
Ann Arbor, MI  48109-1120}
\maketitle
\begin{abstract}
Although no individual piece of experimental evidence for supersymmetry
is compelling so far, several are about as good as they can be with
present errors.  Most important, all pieces of evidence imply the same
values for common parameters --- a necessary condition, and one unlikely
to hold if the hints from data are misleading.  The parameters are
sparticle or soft-breaking masses and $\tan\beta.$ For the parameter
ranges reported here, there are so far no signals that should have
occurred but did not.  Given those parameters a number of predictions
can test whether the evidence is real.  It turns out that the
predictions are mostly different from the conventional \ss ones, and
might have been difficult to recognize as signals of superpartners.
They are testable at LEP2, where neutralinos and charginos will appear
mainly as $\gamma\gamma +$ large $\slashchar{E}$ events, $\gamma +$ very
large $\slashchar{E}$ events, and very soft lepton pairs of same or
mixed flavor.  The results demonstrate that we understand a lot about
how to extract an effective SUSY Lagrangian from limited data, and that
we can reasonably hope to learn about the theory near the Planck scale
from the data at the electroweak scale.
\end{abstract}
\newpage

\noindent{\bf Introduction}

Supersymmetry can allow a solution of the hierarchy problem,
unification of the \sm\ forces, unification of the
\sm\ forces with gravity, provide a derivation of the Higgs mechanism
(which led to the prediction that $M_t$ would be large), and provide a
cold dark matter candidate, the lightest superpartner (LSP).  If string
theory is relevant to understanding weak scale physics in detail it
probably implies supersymmetry at the weak scale.

In this talk I will not emphasize these kinds of evidence for
supersymmetry in nature.  Rather I want to focus on more explicit hints
of effects of superpartners.  What would be nice, of course, is a clear,
explicit, unambiguous effect.  But a little reflection implies that such
an obvious effect is unlikely.  There are two ways effects of \ss can
appear.  Superpartners can be pair-produced as energy or luminosity is
increased at LEP or the Tevatron colliders.  Once the energy threshold
is crossed, luminosity is the important consideration.  That necessarily
means that events will initially appear in small numbers.  Further,
because every superpartner will decay into \sm\ particles plus an
escaping LSP, and \sps will be pair-produced, no event will show a high
resolution mass peak or be uniquely identifiable, as $Z\to \ell^+\ell^-$
was or even $W^\pm \to \ell^\pm \nu$ with only one escaping particle.  A
cursory study of existing limits shows that we would have been lucky to
have observed any \sps so far.  Most published limits depend on extra
assumptions, so generally valid limits are even fewer than reported
ones.

The second way effects could appear is as loop contributions in rare
decays or as small radiative corrections to branching ratios.  The most
likely place for an effect has been known\cite{ferrera} for over two
decades to be $BR(b \to s\gamma)$ both because there is no tree level
contribution so the superpartner loop can be of the same order as the
\sm\ loop, and because in the \sc limit the superpartner and \sm\
contributions must cancel coherently to give a vanishing decay.  But if
the \sc effect is (say) $\sim$ 30\% (quite large), to be statistically
significant the errors have to be $\lsim$ 10\%.  The theoretical error
in the \sm\ value has recently\cite{misiak} decreased to this level
after much difficult effort, so finally an effect here will eventually
be possible to observe.  The present experimental error is about 20\% of
the \sm\ value so that has to decrease too; new data will be reported
eventually by the CLEO collaboration starting spring or summer of 1997.

The other place where \sc loop effects were predicted to be observable
was $R_b = BR(Z\to b \bar b)/BR(Z\to\ \hbox{hadrons}).$ The recent
history here has been complicated.  A few years ago a large deviation
was reported at LEP, leading to renewed theoretical study --- the
calculations are complicated.  The first theoretical studies showed that
effects almost as large as the reported deviations from the \sm\ (then
$\sim$ 2\%) could be obtained.  Then constraints from other data were
put in\cite{pokorski}, and the possible size of the theoretical effect
decreased to a maximum of about 1\%, with a typical value of about
$0.65\pm 0.2$\%.  At the same time reevaluations of the experimental
effect gave a decreasing one\cite{steinberger}. The current world
average is $0.2178 \pm 0.0011$ compared to a \sm\ value of 0.2158, while
the average of four measurements reported in the past year is $.2165 \pm
0.0012$, and the ALEPH group has recently reported $0.2159 \pm 0.0014$,
OPAL $.2178 \pm 0.0022$, and DELPHI $0.2179 \pm 0.0039$.  The
experiments are very difficult.  The bottom line is that by itself $R_b$
cannot be strong evidence for or against a significant \sc effect from
the constrained theory of order ${2\over 3}$\% since the experimental
$1\sigma$ error is of that order.

Fortunately, as we will see below, \ss affects $BR(b\to s\gamma)$ and
$R_b$ in a coordinated way, so when they are combined the $R_b$ data can
still play a useful role.  There is an additional test from the value of
$\alpha_s$.  But the net effect cannot be compelling.

\begin{figure}
\centering
\epsfxsize=13cm
\hspace*{0in}
\epsffile{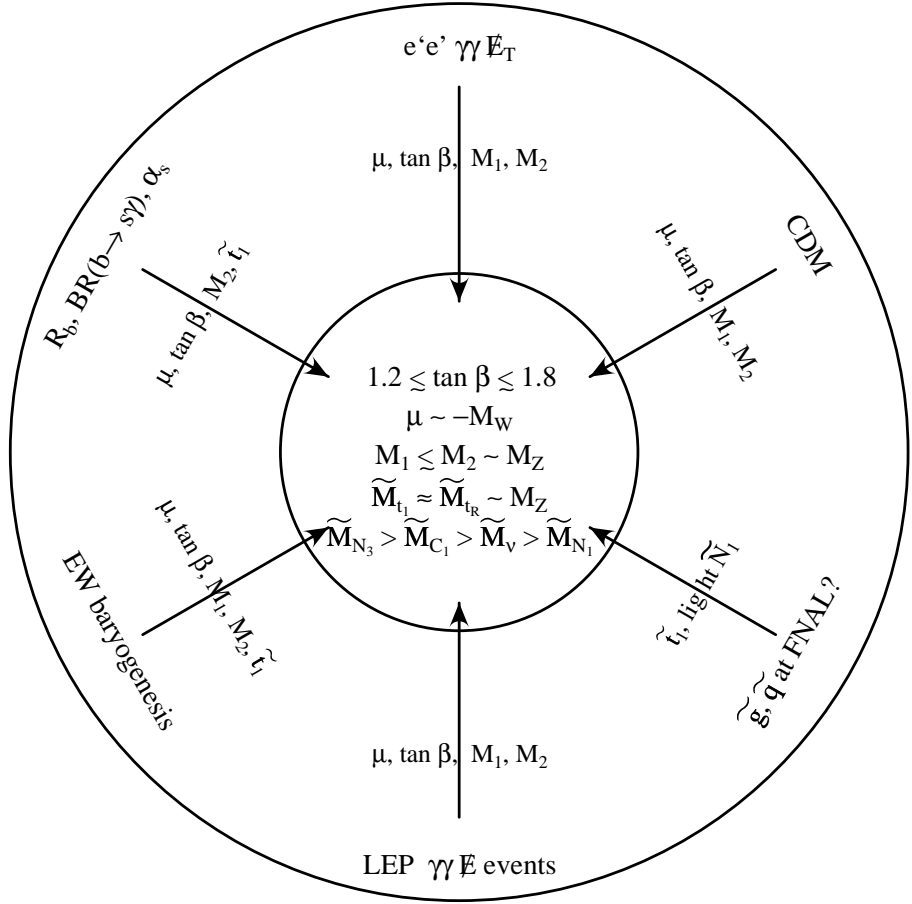}
\caption{Figure 1 shows the existing clues that hint at evidence for
superpartners.  None are compelling, though each is about as strong as
it could be given present errors and integrated luminosities.  If they
were not evidence for superpartners, each might have faked such
evidence, but it is extremely unlikely all would have worked
\underbar{and} given the same result for $\tan\beta$, $\mu$, $M_1$,
$M_2$, $\tilde t_1$.  Each is described in the following sections.  The
region of SUSY parameter space in the inner circle is consistent with
having the indicated phenomena as signals, and with all collider and
decay constraints.  The LSP is mainly higgsino with mass about 50 GeV.}
\end{figure}

In the past two years several weak hints of \ss have emerged, and I will
briefly describe them below.  Each hint can only be interpreted as
evidence for \ss if certain parameters take on certain values.  The
parameters we need to discuss these issues are $\mu$ (which can be
thought of as an effective higgsino mass, $-M_{\rm Planck}$ $\lsim \mu
\lsim M_{\rm Planck}$), $M_1$ and $M_2$ ($U(1)$ and $SU(2)$ gaugino
masses $(0\lsim M_1, M_2 \lsim {\rm TeV})$), $\tan\beta$ (ratio of the
two vacuum expectation values, $1 \lsim \tan\beta \lsim 70$), $\mtildet$
(mass of the light stop mass eigenstate), and $\theta_{\tilde t}$ (a
rotation angle from symmetry eigenstates to mass eigenstates that
measures how much of $\tilde t_1$ is $\tilde t_R$ or $\tilde t_L$, the
\sps of $t_R$ or $t_L$).  The full theory has many more parameters that
will come into play after more \sps are observed, but these few are all
we need for the present.  These parameters determine the masses and
coupling of neutralinos and charginos.  The notation here is that
$\tilde N_i$ represent the four neutralino mass eigenstates, linear
combinations of the \sps of $\tilde \gamma,$ $\tilde Z,$ $\tilde h_U$,
$\tilde h_D$, and $\tilde C^\pm_i$ the two chargino mass eigenstates,
linear combinations of the \sps $\tilde W^\pm$ and $\tilde H^\pm.$
$\tilde N_1$ is the LSP.

If we were being fooled by fluctuations in data, and the various hints
were not actually evidence for supersymmetry, we would expect that
measurements of $\mu$ or $\tan\beta$ or other parameters would give one
value from one bit of evidence, a second value from another bit, and so
on.  Since the allowed ranges are large it would be surprising if
different (misleading) data gave similar values.  What is exciting is
that all the evidence leads to a common set of values for $\mu$,
$\tan\beta$, $M_1$, $M_2$, and $\tilde t_1$!  It is this emergence of a
common set of parameters that is the strongest evidence for \ss today.
It is reminiscent of the testing of the \sm\ by checking whether
different experiments gave a common value of $\sin^2\theta_W$.  Figure 1
summarizes the parameters implied by taking the SUSY clues seriously,
and gives a set of \lq\lq models'' that are consistent with all reported
data.  (The word \lq\lq models'' is used loosely both for the general
class studied here, with a neutralino LSP that is mainly higgsino, and
for particular correlated sets of parameters in the ranges described in
Figure 1.)  Once we have a common set of parameters we can make a number
of predictions.  We will see that several are non-standard SUSY
predictions, mainly for LEP.

The entire analysis described here is done with an effective Lagrangian
at the EW scale, the most general softly broken supersymmetric
Lagrangian.  No unification assumptions are made for soft-breaking
masses, and no assumptions about SUSY breaking.  The data force the
conclusion that the LSP is a mainly -- higgsino neutralino of mass about
50 GeV.  

Of course, if superpartners are indeed being detected, it is very
important for the development of particle physics and astrophysics and
cosmology.  It is also very important for more mundane, practical
reasons -- all of the planning and studies and panels for future
utilization and development and construction of experimental facilities
for particle physics is effectively based on the assumption that no
major discoveries will be made at LEP or FNAL.  If Higgs bosons and/or
superpartners are found at LEP or FNAL, then those facilities will be
able to study them if resources are put into luminosity and detectors
and perhaps small marginal energy increases at LEP.  Much more energetic
facilities may also be of value, but how valuable they are depends on
what is found.

\vspace{.5cm}
\noindent{\boldmath$R_{\bf b}, BR(b\to s\gamma),$ {\bf and}
\boldmath$\alpha_s$}

\begin{figure}[t]
\centering
\epsfxsize=10cm
\hspace*{0in}
\epsffile{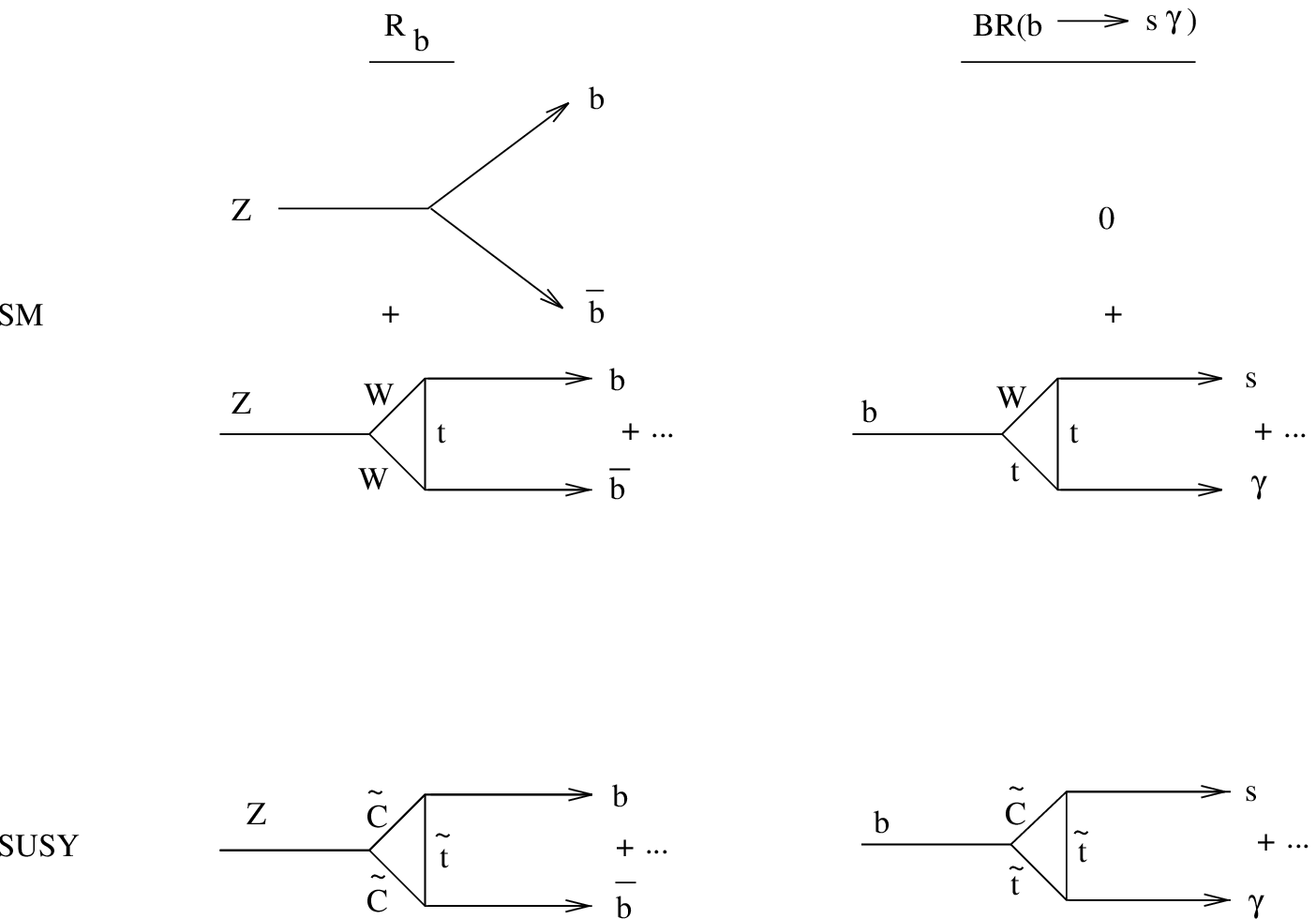}
\begin{center}
{Figure 2}
\end{center}
\end{figure}

As Fig. 2 illustrates, in a \sc theory the two processes $R_b$ and $b\to
s\gamma$ are related because the same \sps occur in loops.  In
particular, as described in the introduction, if \ss is not too broken,
as is the case for the models of figure 1, then $BR(b\!\to\!  s\gamma)$
will be smaller than its \sm\ value.  In that case, as shown in figure
3, $R_b$ must be somewhat larger than its \sm\ value for these models.
(The calculations used to determine the enclosed region in figure 3 are
based on work in progress with M. Carena, C. Wagner, G. Kribs, and
S. Ambrosanio.)  Although neither $R_b$ nor $b\!\to\! s\gamma$ gives a
significant deviation for the \sm, the combined effect is more
significant.  Figure 3 shows both the world average and the past year's
data for $R_b$.

\begin{figure}[h]
\centering
\epsfxsize=10cm
\hspace*{0in}
\epsffile{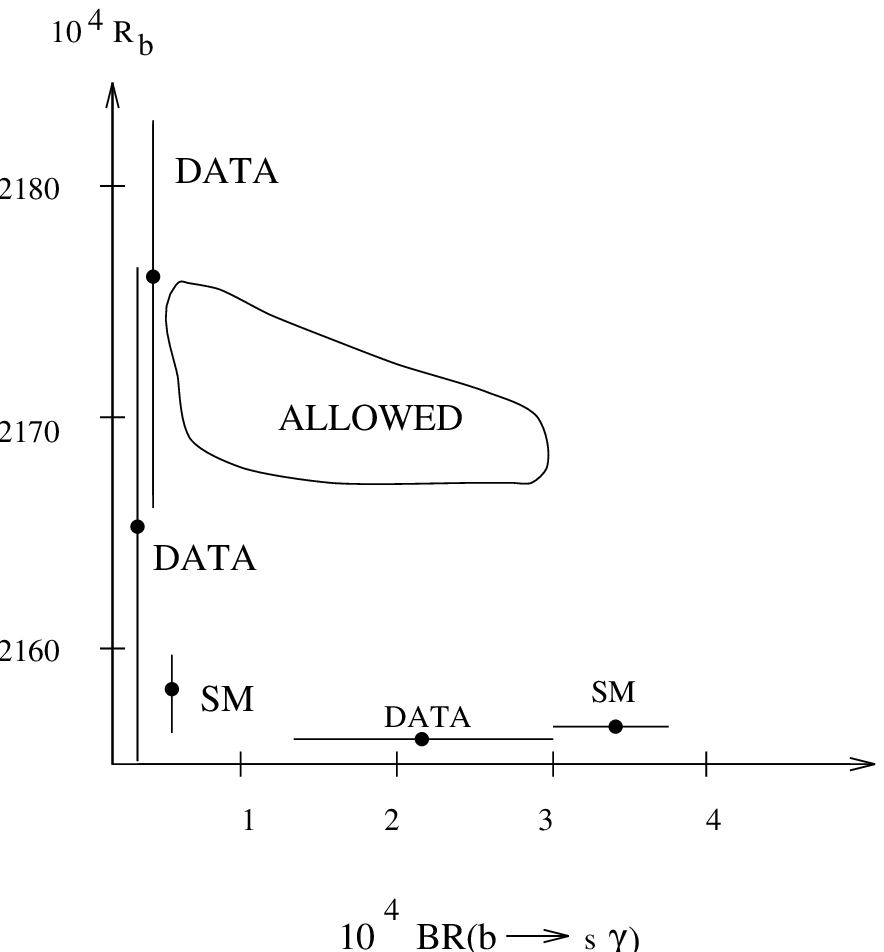}
\begin{center}
{Figure 3}
\end{center}
\end{figure}

Further, there is a constraint that must be satisfied.  If $R_b$ is in
fact increased by the chargino-stop loop, but if that effect is not
taken into account when $\alpha_s(Z)$ is deduced from LEP data, then the
resulting $\alpha_s (Z)$ from $\Gamma_Z$ will be somewhat larger than
the true $\alpha_s$.  Quantitatively, $\delta \alpha_s^{\Gamma_Z} \cong
4 \delta R_B.$ If $\delta R_B \cong 0.0012$, say, then $\delta \alpha_s
\cong 0.005$.  Presently\cite{blondel} $\alpha_s^{\Gamma_Z} = 0.121 \pm
0.003,$ while the world average without the $\Gamma_Z$ value is $0.117
\pm 0.003,$ so a shift of $0.005$ is very consistent.  If this had
failed the whole picture would have been wrong (e.g. if the world
average was above $\alpha^{\Gamma_Z}_s$).  I don't know how to
assign a quantitative measure to the comparison of SUSY and the \sm\
here, but the combined effect of the three observables is certainly of
some interest.

\vspace{.5cm}
\noindent {\bf The CDF e\lq\lq e''\boldmath$\gamma\gamma
\slashchar{E}_T$ Event}

The reported\cite{park} CDF event is interesting for four reasons.  (1)
Most important,\cite{sandro} the probability of the \sm\ giving such an
event is, as far as is known, extremely small, less than $10^{-4}$ from
naive use of the theory and probably considerably smaller when
experimental considerations are included.  While one event can never be
a convincing signal, it is important to understand that this event is
interesting because it should not have occurred in the \sm .  (2) The
event has $\sim 50$ GeV $\slashchar{E_T}$, as expected for SUSY events.
(3) It also has hard isolated $\gamma$'s, predicted long ago\cite{haber}
as one likely way to detect superpartners.  (4) A number of conditions,
cross sections, branching ratios etc. have to come out right or such an
event could not qualify as a SUSY candidate.  It is not easy to satisfy
the conditions.\cite{sandro}

Here I want to emphasize that one way\cite{sandro} to get the photons,
and to interpret this event as production and decay of superpartners, is
for the LSP to be mainly higgsino, and for the next-to-LSP to be mainly
photino.  That interpretation implies the parameters of figure 1, in a
way completely independent of the $R_b, b\!\to\! s\gamma, \alpha_s$
data.  It only uses general features of the CDF event, the presence of
energetic isolated $\gamma$'s and $\slashchar{E}_T$.

\vspace{.5cm}
\noindent{\bf Electroweak Baryogenesis}

Increasingly detailed calculations have been done to determine if the
baryon assymetry of the universe can be generated at the electroweak
scale during the electroweak phase transition.  Recent work\cite{carena}
has concluded that the answer is quantitatively \lq\lq yes'' if
charginos and stops have masses of order the EW scale, and in particular
if $\tan\beta$ is near 1, $\mu <0$, $|\mu| \sim M_1 \sim M_2,$
$M_{h^\circ} \lsim 80$ GeV, and there is a light, mainly right-handed
stop.  These are just the same parameters arrived at by the other
phenomena we consider, and shown in figure 1.

\vspace{.5cm}
\noindent{\bf LEP \boldmath$\gamma\gamma\slashchar{E}$ Events}

Combining LEP data from 161 and 172 GeV, and four detectors, about six
events\cite{wilson} have been reported for $e^+e^- \to \gamma\gamma +$
nothing, with missing invariant mass $\slashchar{M}$ in the region above
$M_Z + 10$ GeV.  We can view the occurrence of such events as a
prediction of the higgsino -- LSP picture suggested by the presence of
the photons in the CDF event, or equivalently as a prediction of the
$R_b + b \to s\gamma + \alpha_s$ data.  In either case\cite{kane} the
events come as $e^+e^- \to \widetilde N_2 (\to \gamma \widetilde N_1)
\widetilde N_2 (\to \gamma \widetilde N_1).$ There is
background\cite{ambrosanio} for such events from $e^+e^- \to \nu\bar\nu$
with two radiated $\gamma$'s.  Assuming $M_{\widetilde N_2} -
M_{\widetilde N_1} > $ 20 GeV to ensure that energetic photons are
likely at FNAL, the signal has $E_\gamma \gsim 8$ GeV while the
background peaks at $E_\gamma \to 0.$ If $M_{\widetilde N_2} -
M_{\widetilde N_1}$ is allowed to decrease to (say) 15 GeV, the minimum
$E_\gamma$ decreases to about 5 GeV.  By making a cut $E_\gamma \gsim 8$
GeV the background cross section is about 1.3 events in the region above
$M_Z + 10,$ where the entire signal should occur since $\slashchar{M} >
2 M_{\widetilde N_1} \cong 100$ Gev.  If we turned the analysis around
and asked what would be required if about 6 such events with 1.3
expected constituted a signal, we would find again $\tan\beta$ near 1,
$\mu < 0,$ $| \mu | \lsim M_1 \sim M_2 \sim M_Z,$ as shown in figure 1.
For LEP184 with $\gsim 50$pb$^{-1}$ per detector, the $E_\gamma$ cut can
be increased to reduce background relative to signal.

\vspace{.5cm}
\noindent{\bf Gluinos and Squarks at the Tevatron?}

It has been argued\cite{mrenna} that an interpretation of the Fermilab
data based on the assumptions that $M_{\tilde g} \gsim M_t + M_{\tilde
t}$ and $M_{\tilde q} \gsim M_{\tilde g}$, with $\tilde g$ and $\tilde
q$ otherwise as light as possible, is at least as consistent as the \sm\
interpretation and perhaps more so in describing reported top-quark data
from Fermilab.  Some tops arise from $\tilde g\to t + \tilde t,$ the
dominant gluino decay since it is the only $2$-body one, and some tops
decay into stops, $t\to \tilde t + \widetilde N_{1,2}.$ If a light stop
exists such an extra source of production of tops may be needed for a
consistent set of top production and decay data.  The total cross
section for $\tilde g + \tilde q$ production $(\tilde g \tilde g ,
\tilde g \tilde q , \tilde q \bar{\tilde q} , \tilde q \tilde q )$ is
about 5 pb, about the same as the Standard Model $t \bar t$ cross
section for 175 GeV top.  Cross sections for different topologies
(dilepton events, $W+$ jets events, and six jet events) will be affected
differently and will give different values.  The total top cross section
will exceed the \sm\ one unless very tight cuts are imposed.  Some
features of the events, such as the $P_T$ of the $t\bar t$ pair, will
behave differently.  All of these phenomena are at least as consistent
with the SUSY predictions as with the SM.  If this is happening it
requires a stop and an $N_1$ that are not very heavy, consistent with
the values in figure 1.

\vspace{.5cm}
\noindent{\bf Cold Dark Matter}

The LSP resulting from the above analyses has well determined
properties.  It is a candidate for the cold dark matter of the universe.
If the CDF event or the LEP $\gamma\gamma\slashchar{E}$ events actually
are evidence of superpartners, then the LSP has effectively been
observed in the laboratory.  It is mainly a higgsino, approximately the
\sp of the Higgs boson.  Before we can conclude it is providing much of
the cold dark matter, we must calculate its relic density.  It could
overclose the universe, in which case the whole picture presented here
would be wrong, or it could provide a negligible amount of CDM because
it annihilates too efficiently.  In fact, for the parameters of figure 1
(it depends on $\tan\beta, \mu , M_1 , M_2$) calculations
give\cite{wells} a relic density just about right to provide a flat
universe with $\sim 2/3$ CDM ($\Omega_{CDM} h^2 \sim 1/4$ to a factor of
two or so).  We could again turn it around, insist that the parameters
be such as to make the LSP a good CDM candidate quantitatively (since
supersymmetry provides such a candidate, which has been known for almost
two decades, surely we do not want to give up that opportunity).  Then
the solution is not unique, but if we insist on a CDM candidate and also
that any of the CDF event, or LEP events, or $R_b + b\!\to\! s\gamma +
\alpha_s$, or electroweak baryogenesis are real effects of supersymmetry
then the solution is uniquely the parameters of figure 1.

\vspace{.5cm}
\noindent{\bf Light Higgs Boson}

In a \sc world the mass of the lightest Higgs boson can be at most about
150 GeV.  Present analyses give $m_{h^\circ} = 117 \pm {107 \atop 64}$
or $m_{h^\circ} < 375$ GeV at 95\% CL from the LEP working
group,\cite{blondel} and $m_{h^\circ} = 101.0 \pm {99.7 \atop 50.2}$ or
$m_{h^\circ} < 312$ GeV at 95\% CL from the analysis\cite{degrassi} of
Degrassi, Gambino, and Sirlin, which includes $\alpha^2 m_t^2$
contributions that are not yet in the LEP working group treatment.  Thus
there is finally statistically significant evidence for a light Higgs
boson, a necessary condition for SUSY to hold.  At this level of
numerical evidence there are no implications for SUSY parameters.
However, the tree-level $m_h$ has an upper limit of $M_Z |\cos 2 \beta
|$ and that is suppressed with $\tan\beta$ near one.  Also, the top loop
radiative correction to $m_h$ is suppressed with a light $\tilde t_R.$
Thus the parameters of figure 1 imply a relatively light $m_{h^\circ}$,
in the range below about $M_Z$ and more likely in the smaller part of
the range.  That $h^\circ$ couples rather like a Standard Model $h$, and
is likely to be observed at LEP2.  It will certainly be observed at FNAL
if it is a little heavy for LEP2.

\vspace{.5cm}
\noindent{\bf Consistency With Other Data, and LEP184 Predictions}

With the parameters in the figure 1 range we can ask what other
experimental predictions can provide tests.  As soon as the masses of
neutralinos and charginos are in the kinematical range allowed by LEP
the production cross sections for some channels become large.  In
particular, $e^+e^- \to \widetilde N_1 \widetilde N_3$ is at the pb
level, so dozens of events must have already been produced at LEP if
figure 1 is correct.  They were not observed.  Does that already exclude
this picture?

A little analysis shows that one simple mass ordering immediately
implies\cite{kane} that almost all $\widetilde N_3$ decays are
invisible. So long as
$$M_{\widetilde N_3}\ \gsim\ M_{\widetilde C_1} \ >\ M_{\tilde \nu}\
> \ M_{\widetilde N_1}$$ then all data at LEP and FNAL is consistent
with the results of Figure 1.  This is the unique way to have
$\gamma\gamma\slashchar{E}$ events without many 
$\widetilde N_1 \widetilde N_3$ events. 

For this mass ordering, $\tilde \nu \to \nu\widetilde N_1$ dominantly
(with a small BR for $\tilde \nu \to \nu \widetilde N_2 (\to \gamma
\widetilde N_1)$); $\tilde\nu$ is mainly invisible.  So $e^+e^- \to
\tilde\nu\tilde\nu$ is a large cross section but mainly unobservable.
And $\widetilde N_3 \to \tilde \nu\nu$ dominantly so if $\tilde \nu$ is
invisible so is $\widetilde N_3.$  Thus not only the LSP, but also
$\tilde\nu$ and $\widetilde N_3$ are effectively invisible.

From these channels, particularly from $\widetilde N_2 (\to \gamma N_1)
\widetilde N_3$,
and from radiation of a detected initial $\gamma$, a detectable excess
is predicted to occur for $e^+e^-\to\gamma +$ invisible, with the excess
having a minimum recoil mass $\slashchar{M}$ of at least $M_{N_1} +
M_{\tilde \nu}$ for some events, and at least $2M_{\tilde \nu}$ for
others, i.e. the excess is only at larger missing invariant mass well
over 100 GeV.  About 80\% of the excess is from decays, the rest from
radiated inital hard $\gamma$'s.  One other visible channel will be
$e^+e^- \to \widetilde N_2 (\to \gamma\widetilde N_1) \widetilde N_2
(\to \gamma \widetilde N_1)$, already described above.

Signatures for charginos and stops also become nonstandard; $e^+ e^- \to
\ell^\mp \ell'^\pm$ $\slashchar{M}$, with large $\slashchar{M}$ and
therefore soft leptons, will dominate for charginos $(\widetilde C^\pm
\to \ell^\pm \tilde\nu)$.  Here $\ell , \ell'$ are charged leptons
perhaps of different types. Detailed signatures are described in
ref. 11.  Since the chargino mass is not much larger than the sneutrino
mass, the leptons can be very soft, perhaps only one or two GeV.  There
is no background from $W^+W^-$ for these soft leptons.  Thus charginos
should appear at LEP2 as events with, say, a one-GeV electron and a 2
GeV muon, acoplaner and acolinear, and nothing else.

\vspace{.5cm}
\noindent{\bf Extracting Supersymmetry Physics From Supersymmetric Data}

Even if the results described here do not turn out to correspond to
reality, the analysis represents an existence proof that the combination
of some data and the tightly constrained SUSY theory is powerful enought
to allow us to extract the relevant masses and SUSY Lagrangian
parameters from the data with pretty good accuracy.  This was aided by
the presence of the photons here, but a similar situation will hold
whatever the form the data takes.  Here we have used a very general
softly broken supersymmetric effective Lagrangian to proceed, with very
few assumptions.  Assumptions about soft-breaking parameters are useful
to study the behavior of the theory before there is data, but once there
is data the parameters should be measured and assumptions tested.
Measuring the parameters of the general effective supersymmetric
Lagrangian at the EW scale will be challenging, but fun and doable,
particularly by combining information from different experiments.

\vspace{.5cm}
\noindent{\bf Implications for Theory}

If we are not being misled by the interesting but not individually
compelling experimental hints of physics beyond the \sm, and by the fact
that they all imply the same set of SUSY parameters --- which would be
remarkable if they were just fluctuations or systematic errors --- then
the results of figure 1 may provide significant information about the
form of the effective Lagrangian at the unification scale (GUT or string
unification).  $\tan\beta$ may be\cite{sandro,carena} very near or even
below its perturbative lower limit, which could provide information
about intermediate scale matter.  $M_1$ may be about equal to $M_2$
rather than the naive prediction $M_1 = {5\over 3} \tan^2 \theta_W M_2
\cong {1\over 2} M_2,$ which would provide important information about
soft-breaking terms[7,9].  $M_{\tilde e_L}$ may be less than $M_{\tilde
e_R}$, which could\cite{kolda} come from D-terms associated with a new
$U(1)$ symmetry, and help determine its charges.  In the MSSM with \sps
below about a TeV it is necessary\cite{pierce} that $\alpha_s(Z)$ be
about 0.13, certainly larger than 0.125.  If $\alpha_s (Z) = 0.117$ from
experiment, new effects must reduce the predicted $\alpha_s$.  However
the results finally come out, we can be optimistic that we will be able
to learn a great deal about the form of the theory near the Planck scale
from data at the EW scale combined with relevant theory.

\vspace{.5cm}
\noindent  {\bf Acknowledgements}

I am very grateful to a number of colleagues for discussions and
collaborations: S. Ambrosanio, M. Carena, H. Frisch, T. Gherghetta,
C. Kolda, G. Kribs, G. Mahlon, S. Martin, S. Mrenna, C. Wagner,
J. Wells, G. Wilson.  This research was supported in part by the
U.S. Department of Energy.


\baselineskip=13pt
\begin{thebibliography}{99}
\bibitem{ferrera}
S.Ferrera and E. Remiddi, Phys. Lett. {\bf 53B} (1974) 347.
\bibitem{misiak}
K. Chetyrkin, M. Misiak, and M. M\"unz, hep-ph/96121313; C. Greub and
T. Hurth, hep-ph/9703349; Z. Ligeti, L. Randall, and M. Wise,
hep-ph/9702322.
\bibitem{pokorski}
See S. Pokorski, Rapporteur's talk at the 1996 International Conference
on High Energy Physics, Warsaw, Aug. 1996.
\bibitem{steinberger}
See the talk of J. Steinberger, XXXI Rencontres de Moriond, Les Arcs,
France, March 1997.
\bibitem{blondel}
A. Blondel, XXXI Rencontres de Moriond, Les Arcs, France, March 1997.
\bibitem{park}
S. Park, \lq\lq Search for New Phenomena in CDF'' 10$^{\rm th}$ Topical
Workshop on Proton-Antiproton Collider Physics, edited by R. Raja and
J. Yoh, AIP Press, 1996.
\bibitem{sandro}
S. Ambrosanio, G.L. Kane, G.D. Kribs, S.P. Martin, and S. Mrenna,
Phys. Rev. Lett. {\bf 76} (1996) 3498.
\bibitem{haber}
H. Haber, G.L. Kane, and M. Quiros, Phys. Lett. {\bf 160B} (1985) 297;
H. Komatsu and J. Kubo, Phys. Lett. {\bf 157B} (1985) 90.
\bibitem{carena}
See M. Carena, M. Quiros, A. Riotto, I. Vilja, C.E.M. Wagner,
hep-ph/9702409 and their references to earlier work on which these
calculations are based.
\bibitem{wilson}
See the talk of G. Wilson, XXXI Rencontres de Moriond, Les Arcs, France,
March 1997.
\bibitem{kane}
G.L. Kane and G. Mahlon, hep-ph/9704450.
\bibitem{ambrosanio}
See S. Ambrosanio,\hfill\break
http://feynman.physics.lsa.umich.edu/ambros/Phys/2Photon+Emiss/; see
also S. Mrenna, \lq\lq Estimating Two Photon + Missing Energy
Backgrounds to SUSY Signals at LEPII'', in preparation.
\bibitem{mrenna}
G.L. Kane and S. Mrenna, Phys. Rev. Lett. {\bf 77} (1996) 3502.
\bibitem{wells}
J. Wells and G.L. Kane, Phys. Rev. Lett. {\bf 76} (1996) 4458.
\bibitem{degrassi}
G. Degrassi, P. Gambino and A. Sirlin, hep-ph/9611363; A. Sirlin,
private communication.
\bibitem{kolda}
C. Kolda and S.P. Martin, Phys. Rev. {\bf D53} (1996) 3871.
\bibitem{pierce}
See the review of D. Pierce, hep-ph/9701344.
\end{thebibliography}
\end{document}